\documentclass[%
 reprint,
showpacs,
 amsmath,amssymb,
 aps,
 prl,
]{revtex4-1}

\usepackage{graphicx}
\usepackage{dcolumn}
\usepackage{bm}
\usepackage{color}

\begin{document}

\preprint{APS/123-QED}

\title{A local fluctuation theorem for large systems}

\author{Guillaume Michel$ ^{1,2} $}
\email{email: guillaume.michel@ens.fr}
\author{Debra J. Searles$^{2,3}$}%
\email{email: d.bernhardt@uq.edu.au}
 \affiliation{%
$^{1}$D\'epartement de Physique, \'Ecole Normale Sup\'erieure, 24 Rue Lhomond, 75 005 Paris, France \\ 
$^{2}$Queensland Micro- and Nanotechnology Centre,
Griffith University, Brisbane Qld 4111, Australia\\
$^{3}$Australian Institute of Bioengineering and Nanotechnology and School of Chemistry and Molecular Biosciences,
The University of Queensland, Brisbane Qld 4072, Australia}%
\date{\today}
\begin{abstract}
The fluctuation theorem characterizes the distribution of the dissipation in nonequilibrium systems and proves that the average dissipation will be positive. For a large system with no external source of fluctuation, fluctuations in properties will become unobservable and details of the fluctuation theorem are unable to be explored.  In this letter, we consider such a situation and show how a fluctuation theorem can be obtained for a small open subsystem within the large system. We find that a correction term has to be added to the large system fluctuation theorem due to correlation of the subsystem with the surroundings. Its analytic expression can be derived provided some general assumptions are fulfilled, and its relevance it checked using numerical simulations.
\end{abstract}

\maketitle

\section*{Introduction}
In classical thermodynamics, nonequilibrium systems are irreversible.  That is, the second law of thermodynamics stipulates that a macroscopic system evolves in one direction and fluxes have a definite sign.   However, the second law is strictly applicable only to large systems or where properties are measured over long time scales. As the size of the system decreases, unusual events caused by thermal fluctuations become more frequent, and average values of the flux with the opposite sign to that predicted for a thermodynamic system are observable over finite periods. These effects are quantified by the Fluctuation Theorem (FT), which states that for a system in a known initial state and driven out of that state,
\begin{equation} \label{eq:TFT}
\ln \left( \frac{\mathrm{p}(\Omega_{t} =A)}{\mathrm{p}(\Omega_{t} =-A)} \right)=A
\end{equation}
where $\Omega_{t}$ is an extensive function, the dissipation function, measured over a period of $t$ \cite{evans_fluctuation_2002}, which describes how irreversible a process is. It is formally defined as $\Omega_{t}=\ln (f(\Gamma)/f(S^t\Gamma))-\Lambda_t$ where $f$ is the initial phase space distribution function, $\Gamma$ is a point in phase space, $S^t$ is the time evolution operator and $\Lambda_t$ is the phase space expansion over the period, $t$. In (\ref{eq:TFT}), $\mathrm{p}(\Omega_{t} =A)$ refers to the probability that $A$ takes on a value $A\pm dA$.  Eq. (\ref{eq:TFT}) is also asymptotically valid for steady-states. Initially based on a heuristic derivation and numerical evidence \cite{Evans_1993}, fluctuation relations have been derived for closed systems in many different frameworks (see \cite{NEMD_preuveEvans,evans_fluctuation_2002,NEMD_preuveGallavotti,*NEMD_preuveGallavotti2,Markov} for early results).  Recently the fluctuation theorem has been used to measure  physical quantities in experiments, for instance the torque of a molecular motor \cite{m_motor}. In general, the use of the FT is relevant when the work done by the external forces is similar to the thermal energy $k_BT$ (or its equivalent if another kind of thermal equilibrium is reached, \textit{e.g.} \cite{SSFT_Granular}). Otherwise, it reduces to stating irreversibility, because the probability of observing a negative dissipation is so small that it can be considered impossible. Therefore, in large systems, (\ref{eq:TFT}) can not be properly tested or applied. To deal with such a situation, one can record the dissipation over a small open subsystem where thermal fluctuations are sizeable. The dissipation in the subsystem will not satisfy equation (\ref{eq:TFT}) in general, and in this manuscript we consider if an analytical expression for a Local Fluctuation Theorem (LFT) can be obtained.
\\ \indent  Some experimental and theoretical work has been carried out on local fluctuation theorems since 1998.  Measurements of local properties for steady states have been shown to satisfy fluctuation relations \cite{LFT_Ciliberto,LFT_Ciliberto2,SSFT_Granular,LFT_Shang}. However, these studies employ an effective temperature, which can be considered to provide an \textit{ad hoc} correction coefficient to the fluctuation relations. Gallavotti showed why the fluctuation theorem is expected to be valid at long times for the local entropy creation rate for a class of weakly coupled systems \cite{LFT_Gallavotti}. Ayton et al. \cite{LFT_Searles} obtained a local fluctuation theorem for the dissipation function (\ref{eq:TFT}), and provided numerical results. In the present paper, we show that in a highly correlated system, the local dissipation obeys a LFT which is (\ref{eq:TFT}) with a linear correction term.  The result explains the previous effective temperatures, the correction term can be analytically described in some cases, and it vanishes in the low correlation limit.
\\ \indent In the next section we derive a LFT, then we numerically investigate the relevance of our assumptions in a realistic system.
\section*{A Local Fluctuation Theorem}
Consider a large system that fulfils the assumptions of the FT, {\it i.e.} the invariance of the initial distribution function under time-reversal mapping, ergodic consistency and time-reversibility of the equations of motion \cite{evans_fluctuation_2002}. The FT can then be derived for an arbitrary phase function $\Phi(\boldsymbol{\Gamma})$ odd under time-reversal \cite{searles_2007},  
\begin{equation} \label{LFT_1}
\ln \left( \frac{\mathrm{p}(\Phi_{t} =A)}{\mathrm{p}(\Phi_{t} =-A)} \right)= - \ln \left( \langle e^{-\Omega_t}   \rangle_{\Phi_t=A} \right)
\end{equation}
where $\langle ... \rangle_{\Phi_t=A} $ is the conditional ensemble average over points for which $\Phi_t=A \pm dA$.   If $\Phi$ is chosen to be the dissipation function, (\ref{LFT_1}) becomes the usual FT. Here, we split $\Omega$ into two contributions $\Omega_\ell$ and $\Omega_\ell^*$, where $\Omega_\ell$ is the dissipation function measured over an arbitrary volume of length $\ell$, and choose $\Phi=\Omega_\ell$. In this case, (\ref{LFT_1}) becomes a local fluctuation theorem
\begin{equation} \label{LFT_2}
\ln \left( \frac{\mathrm{p}(\Omega_{\ell,t} =A)}{\mathrm{p}(\Omega_{\ell,t} =-A)} \right)= A - \ln \left( \langle e^{-\Omega_{\ell,t}^*}   \rangle_{\Omega_{\ell,t}=A} \right)
\end{equation}
This equation provides an exact expression for the correction term. If there is no correlation between the two local dissipations, this term vanishes and $\Omega_\ell$ obeys a bare FT (\ref{eq:TFT}). Otherwise, some assumptions have to be made to obtain an effective description of this term. We will make two main assumptions. The first one is that one can consider the local dissipation as a random variable instead of a phase function so that, 
\begin{equation} \label{Correlation}
\Omega_{\ell,t}^*=\alpha \Omega_{\ell,t}+\xi
\end{equation}
where $\alpha $ is a coefficient that depends on the shape and size of the subvolume, and $\xi$ a random variable that is not correlated with $\Omega_{\ell,t}$. Physically, the first term describes the very strong correlations that exist near the boundaries between the two domains, and $\xi$ stands for the dissipation far away from the volume of interest, which does not depend on $\Omega_{\ell,t}$. Quadratic and higher order terms are neglected in (\ref{Correlation}), but do not seem relevant. This relation between the two local dissipations leads to an exact expression for the LFT,
\begin{equation} \label{LFT_3}
\ln \left( \frac{\mathrm{p}(\Omega_{\ell,t} =A)}{\mathrm{p}(\Omega_{\ell,t} =-A)} \right)= (1+\alpha)A
\end{equation}
This model may explain why the left hand side of (\ref{LFT_3}), called the asymmetry function, has been experimentally found to be a linear function of A with a slope that is not unity in a number of experiments ({\it e.g.} \cite{SSFT_Granular}). Given our assumption of non-correlation between $\xi$ and $\Omega_{\ell,t}$, the coefficient $\alpha$ is
\begin{equation} \label{alpha} 
\alpha = \dfrac{\langle \Omega_{\ell,t}^* \Omega_{\ell,t} \rangle - \langle \Omega_{\ell,t}^* \rangle \langle \Omega_{\ell,t} \rangle}{\langle \Omega_{\ell,t}^2 \rangle - \langle \Omega_{\ell,t} \rangle^2}.  
\end{equation}
In solid states, diffusion of atoms or molecules is limited and typical lengths of correlation in the dissipation depends on details of the system, the property considered and the field. In some cases they are smaller than the sub-volume that has to be considered for a LFT to be relevant. In other cases, such as systems close to a phase-transition, an analytic expression for $\alpha$ can often be derived (\textit{e.g.} for the Ising model). In fluids, molecules travel thought several subvolumes, so correlation lengths are large, and as we shall see they can be related to diffusion lengths. 
We focus on such systems, consisting a fluid driven out of equilibrium by an external field that produces a dissipative flux.  This is very general and includes the studies on fluidized granular medium \cite{SSFT_Granular} and Rayleigh-B\'enard convection \cite{LFT_Ciliberto}, but can also model Poiseuille or Couette flows, diffusion processes, and so on. To simplify the notation we consider the case where the field, $F_e$ and dissipative flux $J$ are in the same direction and then the dissipation function reads
\begin{equation}
\Omega_{t}=\beta F_e  J_t
\end{equation}
where $\beta$ is the inverse temperature to which the system would relax in the absence of the field, and we adopt the notation that the average value of $J$ is positive in a dissipative system. Then $\alpha$ refers to the spatial correlations of the integrated flux $J_t$~:
\begin{equation} \label{alphaflux} 
\alpha = \dfrac{\langle J_{\ell,t}^* J_{\ell,t} \rangle - \langle J_{\ell,t}^* \rangle \langle J_{\ell,t} \rangle}{\langle J_{\ell,t}^2 \rangle - \langle J_{\ell,t} \rangle^2}  
\end{equation}
Another expression for this coefficient can be derived after defining details about the subvolume. We consider a rectangular unit cell of length $\mathcal{L}$ and with field applied in the $x$-direction. The subvolumes are obtained by dividing the cell into slices orthogonal to the direction of the field with width $\ell$ (see figure \ref{channel}), but note that the following computations can be adapted to other situations. In this case, the relevant correlations are fully described by the function $C(x)$ defined by
\begin{equation} \label{defL}
C(x)=\dfrac{\langle j_{t}(0) j_{t}(x) \rangle - \langle j_{t}(0) \rangle \langle j_{t}(x) \rangle} {\langle j_{t}(0)^2\rangle - \langle j_{t}(0) \rangle^2}
\end{equation}
where for $J_{\ell,t}$ centred on $x$, $j_t(x) = \lim_{\ell \to 0}J_{\ell,t}/\ell$ is the flux density at $x$. This function describes the decay of spatial correlations~: if there is no correlation between the flux in these two sections, as when $x$ goes to infinity, this term is equal to zero. On the other hand, $C(0)=1$ due to the normalisation. A typical correlation length in the fluid can then be defined by
\begin{equation} \label{l0}
\ell_0=\int_0^\infty C(x) \mathrm{d}x
\end{equation}
For a periodic system, we can replace this with $\ell_0=\int_0^{\mathcal{L}/2} C(x) \mathrm{d}x$ provided the correlations have decayed at $x<\mathcal{L}/2$. This length is to be compared to the subvolume's to know whether a bare fluctuation theorem (\ref{eq:TFT}) is expected to be valid ($\ell \gg \ell_0$) or if a LFT is required. 

We then make a second general assumption~: noting that the function $C$ satisfies $C(0)=1$, $C(\infty)=0$ and has a typical decay length of $\ell_0$, we assume it is modelled by an exponential decay $C(x)=e^{-x/\ell_0}$ where $x\ge0$. Considering (\ref{defL}) this implies that $\langle j_{t}(0) j_{t}(x) \rangle - \langle j_{t}(0) \rangle \langle j_{t}(x) \rangle=Be^{-x/\ell_0}$ where $B$ is a constant. $\alpha$ is then explicitly computable and assuming $\mathcal{L}$ is large, (\ref{LFT_3}) becomes 

\begin{equation} \label{LFT_final}
\ln \left( \frac{\mathrm{p}(\Omega_{\ell,t} =A)}{\mathrm{p}(\Omega_{\ell,t} =-A)} \right)= (1+\dfrac{1}{\frac{\ell}{\ell_0 (1-e^{- \ell/ \ell_0})} -1})A
\end{equation}
This LFT provides an analytic expression for the correction term, which vanishes in the low correlation limit ($\ell \gg \ell_0$) and can be used instead of an effective temperature.  In derivation of (\ref{LFT_final}), we consider the large system limit, $\mathcal{L}-\ell \gg \ell_0$. 

We now consider a particular case and show that it is possible to derive an expression for the correlation length under some conditions.  Our system consists of $N$ particles of charge $c_i$ subject to a field in the $x$ direction, and the dissipative flux becomes $J_t = \int_0^t \sum_{i=1}^N c_i v_{x,i}\mathrm{d}s$ where $v_{x,i}$ is the $x$ component of the velocity for particle $i$.   For a system close to equilibrium, the correlation length, $\ell_0$ takes its origin in the Brownian motion of the particles, and for large $\mathcal{L}$ the following equalities hold~:
\begin{equation}
\int_0^{\mathcal{L}/2}  \langle j_{t}(0) j_{t}(x) \rangle \mathrm{d}x = \dfrac{\langle (\Delta x)^2 \rangle \sum_{i=1}^N c_i^2}{2\mathcal{L}}
\end{equation}

\begin{equation}
\langle j_{t}(0) j_{t}(0) \rangle  \sim \dfrac{ \langle \mid \Delta x \mid \rangle \sum_{i=1}^N c_i^2}{\mathcal{L}}
\end{equation}
where $\Delta x =x(t)-x(0)$ is the $x$-displacement for one particle and $\mathcal{L}$ the length of the system, which does not appear in the final equilibrium expression for $\ell_0$,
\begin{equation} \label{ell0final}
\ell_0 \sim \dfrac{\langle (\Delta x)^2 \rangle }{2 \langle \mid \Delta x \mid \rangle} = \sqrt{ \dfrac{\pi \langle (\Delta x)^2 \rangle}{8} }
\end{equation}
The last equality comes from the assumption of a Gaussian distribution for $\Delta x$. If the external field is not too high, $\ell_0$ will be close to its equilibrium value. 

Therefore, we have seen that for a fluid (or gas) driven out of equilibrium by an external field, the local dissipation recorded in a section of length $\ell$ fulfils (\ref{LFT_final}) provided~:  the system is large and the decay of correlations is reasonably described by an exponential function.  Moreover, at equilibrium the typical length-scale of correlations in the flux density is given by the diffusion length.  As we shall see, if the field is not too large, $\ell_0$ is close to this value which provides a useful way of determining whether or not a bare fluctuation theorem can be applied without measuring the flux-correlations.
\section{Numerical study}
\begin{figure}[b]
    \begin{center} \includegraphics[width=8.6cm]{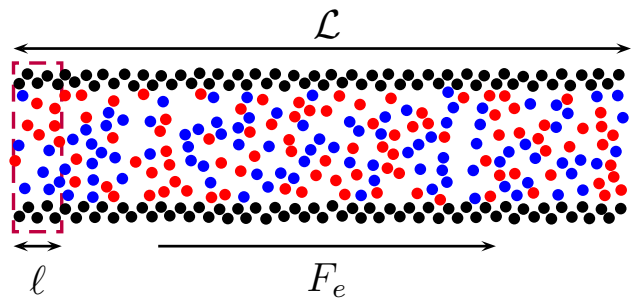} \end{center}
    \vspace*{-0.4cm}
    \caption{A schematic diagram of the system studied. Wall particles are thermostated and represented in black ($c_i=0$), whereas fluid particles can be blue ($c_i=1$) or red ($c_i=-1$).}
    \label{channel}
 \end{figure}

To demonstrate an application of this LFT, check the relevance of our assumptions and verify that the coefficient $\ell_0$ is related to a diffusive length-scale, a system of color-charged particles between atomic walls was numerically studied. This model is one of the first used in nonequilibrium molecular dynamics and is a simplified model of an ionic liquid where the particles do not have Coulomb interactions with each other but experience a force proportional to their color charge when subject to a field. Its simplicity allows fast computation without loss of physical details relevant to the analysis presented in this letter.  The model and its fluctuation relations are discussed in detail in \cite{evans_fluctuation_2002}. The system is a high density gas in a long channel surrounded by thermostated walls. The dynamics are, 
\begin{subequations}
\begin{align} 
& \dot{\textbf{q}}_{i}=\frac{\textbf{p}_{i}}{m}\\
& \dot{\textbf{p}}_{i}=\textbf{F}_{i}(\textbf{q})+c_{i}F_{e}\textbf{e}_{x}-S_i (\alpha \textbf{p}_{i} + k (\textbf{q}-\textbf{q}_{eq}))
\end{align}
\end{subequations}
where $\textbf{q}_{i} $, $\textbf{p}_{i} $ and $c_i = (-1)^i$ are the coordinates, momenta and colour of the $i$th particle ($c_i=0$ for wall particles) and $S_i$ is a switch equal to $1$ for the wall particles and zero otherwise. $\textbf{F}_{i}$ is the interparticle force on a particle, derived from a Weeks-Chandler-Anderson short-ranged repulsive pair potential \cite{WCA}, $k$ the strength of the traps that fix the positions of wall particles, and $F_{e}$ the external field which induces a flux $J = \sum_{i=1}^N c_i v_{x,i}$. Finally, $ \alpha $ is a Gaussian thermostat that fixes the kinetic energy of the walls~: the fluid particles are not thermostated and obey their natural dynamics. The simulation used 320 particles in a two-dimensional space with periodic boundary conditions. The wall temperature was set at 1, the wall density at 0.8, the fluid density at 0.4, the field at $F_e=0.08$, the length of system at $\mathcal{L}=50.6$ and the integration time at $t=80$. All trajectories started from the equilibrium distribution. The volume was divided into subvolumes as described above.  This system is shown in fig.\ref{channel}.

\begin{figure}[tb]
    \begin{center} \includegraphics[width=8.6cm]{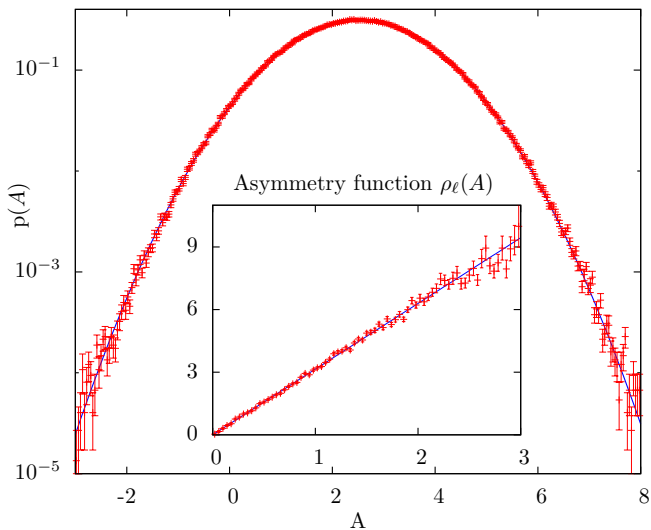} \end{center}
        \vspace*{-0.4cm}
    \caption{PDFs for the local dissipation function in a sub-volume 14.4 times smaller than the total volume and its Gaussian fit. Its mean value is $\langle \Omega _{\ell,t} \rangle \simeq 2.5$ and the asymmetry function, $\rho_\ell(A)=\ln \left(\mathrm{p}(\Omega_{\ell,t} =A)/\mathrm{p}(\Omega_{\ell,t} =-A) \right)$,  is found to be a straight line of slope $3.147\pm 0.001$.}
    \label{pdf_localdissipation}
 \end{figure}

The mean value of the total dissipation is $\langle \Omega_t \rangle \simeq 36$ and in 500 000 samples no negative value was observed. On the other hand, the local dissipation recorded in the volume of length $\ell$ shown in fig.\ref{channel}, was 14.4 times smaller than the total volume.  It consequently produced a number of negative values of  $\Omega_t$ and its asymmetry function can be computed. As expected, it is a linear function with a slope larger than one, see fig. \ref{pdf_localdissipation}.

 \begin{figure}[tb]
    \begin{center} \includegraphics[width=8.6cm]{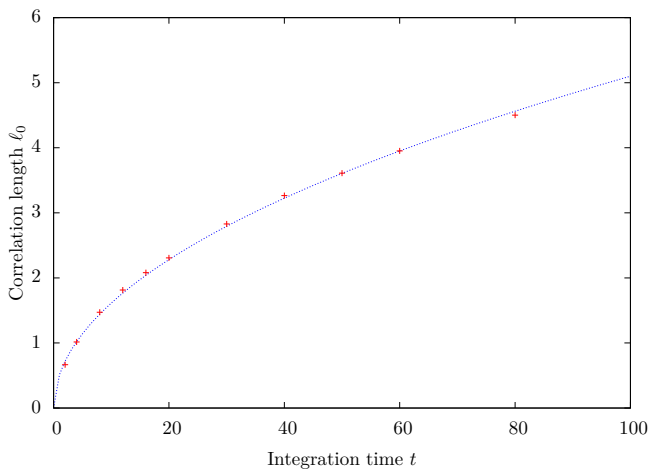} \end{center}
        \vspace*{-0.4cm}
    \caption{Evolution of the correlation length $\ell_0$ determined out of equilibrium using a fit of $C(x)=e^{-x/\ell_0}$ to the simulated data at various times. The dashed line is it fit to a function of $\sqrt{t}$.}
    \label{l0vst}
 \end{figure}
 
\begin{figure}[t]
    \begin{center} \includegraphics[width=8.6cm]{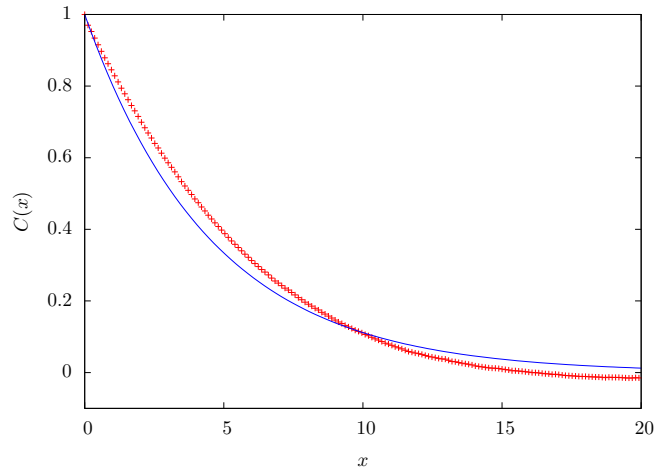} \end{center}
      \vspace*{-0.4cm}
    \caption{The decay of spatial correlations in the integrated flux  calculated with (\ref{defL}) (crosses) and a fit to the exponential model $e^{-x/\ell_0}$, where $\ell_0$ is the correlation length (solid line).}
    \label{fluxcorrelation}
 \end{figure}

The correlation length $\ell_0$ was measured out of equilibrium by fitting $C(x)=e^{-x/\ell_0}$ to the simulated values of $C(x)$ and was found to grow as the square root of the integration time, see fig. \ref{l0vst}.  Close to equilibrium, $\ell_0$ is still expected to be well approximated by a diffusive process, and in this simulation we find $\ell_0=0.6 \sqrt {\langle (\Delta x)^2 \rangle}$, where the equilibrium value of $ \sqrt {\langle (\Delta x)^2 \rangle}$ is obtained using the Einstein relation.  This tends to show that $\ell_0$ would often be well approximated by its equilibrium value in real experiments, as external fields used in molecular dynamics simulations are very large in comparison (in this simulation, the mean velocity eventually reaches a few percent of the thermal one). The main assumption of the derivation was that an exponential decay $e^{-x/\ell_0}$ would fit the function $C(x)$, which is not exact, cf. fig.\ref{fluxcorrelation}.  

\begin{figure}[t]
    \begin{center} \includegraphics[width=8.6cm]{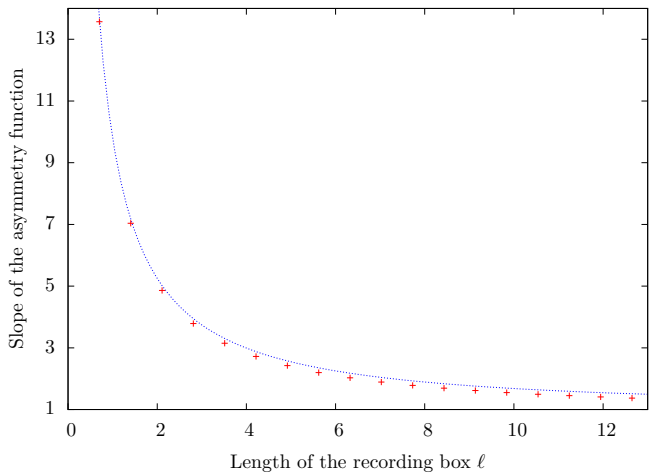} \end{center}
     \vspace*{-0.4cm}
    \caption{Slope of the asymmetry function (left hand side of (\ref{LFT_final})) versus the length of the subsystem. The dashed line is the theoretical result derived in section 2, with the coefficient $\ell_0$ estimated using the data from fig. \ref{l0vst}.}
    \label{slopevst}
 \end{figure}

Using the calculated value of $\ell_0=4.56$ at $t=80$, and $\ell=3.52$, the slope predicted by (\ref{LFT_final}) is $3.30$ which is in good agreement with the numerically determined slope of $3.146 \pm 0.001$ determined from the data shown in fig. \ref{pdf_localdissipation}.  In fig.\ref{slopevst}, the predicted and actual slopes are compared for a range of $\ell$, and are shown to be in very good agreement.  Therefore, even if the correlation decay is not exactly exponential, which could happen as this function is likely to be system-dependant, the predictions of this LFT remain robust.  This result will be verified by application to other systems and in other conditions. 
\section*{Conclusion}
In this letter, we considered a class of  systems (liquid or gas) driven out of equilibrium by an external field. A local fluctuation theorem is derived for the dissipation measured in a section orthogonal to the field and provides an explicit expression for the slope of its asymmetry function. The proof can be adapted to describe other shapes of subvolumes. These results suggest that arbitrary effective temperatures used to account for the effect of measuring a local property can be attributed to a physical effect. This result holds provided some assumptions are fulfilled, mainly a low external field and exponential decay of correlations in the flux. According to the numerical study, this last condition is not restrictive, and a reasonable accordance is enough for this LFT to be verified. Finally, we pointed out that the need to resort to a LFT depends on the ratio $\ell/\ell_0$, where $\ell$ is the length of the subvolume over which the local dissipation is measured and $\ell_0$ is related to an equilibrium diffusion process and can be easily measured via the Einstein relation. As discussed in \cite{searles_2007}, this FT is also asymptotically valid for steady states with an additional assumption of a form of decay of correlations in the dissipation with time.
\\ \indent We would like to thank the Australian Research Council for the support of this project and Dr James C. Reid for helpful discussions.
\end{document}


\preprint{APS/123-QED}

\section*{Supplementary Material}
Here we provide detailed derivations of some of the equations that appear in the manuscript.

The following equations have already appeared in the literature:
\begin{equation} \label{eq:TFT}
\ln \left( \frac{\mathrm{p}(\Omega_{t} =A)}{\mathrm{p}(\Omega_{t} =-A)} \right)=A
\end{equation}

\begin{equation} \label{LFT_1}
\ln \left( \frac{\mathrm{p}(\Phi_{t} =A)}{\mathrm{p}(\Phi_{t} =-A)} \right)= - \ln \left( \langle e^{-\Omega_t}   \rangle_{\Phi_t=A} \right)
\end{equation}

\begin{equation} \label{LFT_2}
\ln \left( \frac{\mathrm{p}(\Omega_{\ell,t} =A)}{\mathrm{p}(\Omega_{\ell,t} =-A)} \right)= A - \ln \left( \langle e^{-\Omega_{\ell,t}^*}   \rangle_{\Omega_{\ell,t}=A}.\right)
\end{equation}

In this work we propose the relationship:
\begin{equation} \label{Correlation}
\Omega_{\ell,t}^*=\alpha \Omega_{\ell,t}+\xi
\end{equation}
where $\alpha$ is a constant that depends on $t$, the size and shape of the volume, and $\xi$ is uncorrelated with $\Omega_{\ell,t}$.  From straightforward substitution of eq. \ref{Correlation}, into eq. \ref{LFT_2} we get:

\begin{equation} \label{LFT_2sup}
\begin{split}
&\ln \left( \frac{\mathrm{p}(\Omega_{\ell,t} =A)}{\mathrm{p}(\Omega_{\ell,t} =-A)} \right) \\
&=A - \ln \left( \langle e^{-\alpha \Omega_{\ell,t}}   \rangle_{\Omega_{\ell,t}=A} \right) - \ln \left( \langle e^{\xi}   \rangle_{\Omega_{\ell,t}=A} \right) \\ 
&=(1+\alpha)A-  \ln \left( \langle e^{\xi}   \rangle \right).
\end{split}
\end{equation}
When $A=0$ we have the trivial relationship $\ln \left( \frac{\mathrm{p}(\Omega_{\ell,t} =A)}{\mathrm{p}(\Omega_{\ell,t} =-A)} \right)=0$, therefore the rhs of the equation \ref{LFT_2sup} must also be zero: $\ln \left( \langle e^{\xi}   \rangle \right)=0$ and we obtain:
\begin{equation} \label{LFT_3}
\ln \left( \frac{\mathrm{p}(\Omega_{\ell,t} =A)}{\mathrm{p}(\Omega_{\ell,t} =-A)} \right)= (1+\alpha)A.
\end{equation}

From \ref{Correlation} we can also write:
\begin{equation}
\langle \Omega_{\ell,t}^* \Omega_{\ell,t} \rangle=\alpha \langle  \Omega_{\ell,t} \Omega_{\ell,t} \rangle+\langle \xi \Omega_{\ell,t} \rangle
\end{equation}
and
\begin{equation}
\langle \Omega_{\ell,t}^* \rangle \langle\Omega_{\ell,t} \rangle= \alpha \langle  \Omega_{\ell,t} \rangle \langle \Omega_{\ell,t} \rangle+\langle \xi \rangle \langle\Omega_{\ell,t} \rangle
\end{equation}
Noting that $\langle \xi \Omega_{\ell,t} \rangle=\langle \xi\rangle \langle \Omega_{\ell,t} \rangle$. These equations can be combined to give:
\begin{equation} \label{alpha} 
\alpha = \dfrac{\langle \Omega_{\ell,t}^* \Omega_{\ell,t} \rangle - \langle \Omega_{\ell,t}^* \rangle \langle \Omega_{\ell,t} \rangle}{\langle \Omega_{\ell,t}^2 \rangle - \langle \Omega_{\ell,t} \rangle^2}  
\end{equation}

Now consider field driven flow where:
\begin{equation}
\Omega_{t}=\beta F_e  J_t
\end{equation}
then
\begin{equation} \label{alphaflux} 
\alpha = \dfrac{\langle J_{\ell,t}^* J_{\ell,t} \rangle - \langle J_{\ell,t}^* \rangle \langle J_{\ell,t} \rangle}{\langle J_{\ell,t}^2 \rangle - \langle J_{\ell,t} \rangle^2}  
\end{equation}
Then defining $\lim_{\ell \to 0} J_{\ell,t}/\ell=j_t(x)$, such that  $ J_{\ell,t}^*=  \int_{-\mathcal{L}/2}^{-\ell/2} j_t(x) dx+\int_{\ell/2}^{\mathcal{L}/2} j_t(x) dx$, $ J_{\ell,t}= \int_{-\ell/2}^{\ell/2} j_t(x) dx$ and $\lim_{\ell \to 0}\int_{-\ell/2}^{\ell/2} (j_t(x)/\ell) dx=j_t(0)$ we can substitute into \ref{alphaflux} to obtain:
\begin{equation} \label{alphaflux_sup1} 
\alpha = \dfrac{2\int_{\ell/2}^{\mathcal{L}/2} \int_{-\ell/2}^{\ell/2} \left( \langle j_t(x_1)j_t(x_2) \rangle -  \langle  j_t(x_1) \rangle \langle j_t(x_2)\rangle\right) dx_2 dx_1 }
{\int_{-\ell/2}^{\ell/2} \int_{-\ell/2}^{\ell/2} \left( \langle j_t(x_1)j_t(x_2) \rangle - \langle j_t(x_1) \rangle\langle j_t(x_2) \rangle \right) dx_2 dx_1}  
\end{equation}
where the symmetry of the correlations in a homogeneous system has been used.  Then
\begin{equation} \label{alphaflux_sup2} 
\lim_{\ell \to 0}\alpha = \dfrac{2\int_0^{\mathcal{L}/2}  \left( \langle j_t(x)j_t(0) \rangle -  \langle  j_t(x) \rangle \langle j_t(0)\rangle\right) dx }{(\langle j_t(0)^2 \rangle - \langle j_t(0) \rangle^2)\ell}  
\end{equation}

It is useful to define,
\begin{equation} \label{defL}
C(x)=\dfrac{\langle j_{t}(0) j_{t}(x) \rangle - \langle j_{t}(0) \rangle \langle j_{t}(x) \rangle} {\langle j_{t}(0)^2\rangle - \langle j_{t}(0) \rangle^2}
\end{equation}
and clearly, $\lim_{\ell \to 0} \alpha=2\int_0^{\mathcal{L}/2} (C(x)/\ell) dx$.  If the correlation between $j(0)$ and $j(x)$ decays within $\mathcal{L}/2$ we can define:

\begin{equation} \label{l0}
\ell_0=\int_0^{\mathcal{L}/2} C(x) \mathrm{d}x
\end{equation}

We now assume that $C(x)=e^{-x/\ell_0}$ where $x>0$.  Considering (\ref{defL}) this implies that $\langle j_{t}(0) j_{t}(x) \rangle - \langle j_{t}(0) \rangle \langle j_{t}(x) \rangle=Be^{-x/\ell_0}$ where $B$ is a constant and $x>0$, or more generally $\langle j_{t}(x_1) j_{t}(x_2) \rangle - \langle j_{t}(x_1) \rangle \langle j_{t}(x_2) \rangle=Be^{-|x_2-x_1|/\ell_0}$.

Now,
\begin{equation} 
\begin{split}
&\int_{-\ell/2}^{\ell/2} \int_{-\ell/2}^{\ell/2} \left( \langle j(x_1)j(x_2) \rangle - \langle j(x_1) \rangle\langle j(x_2) \rangle \right) dx_2 dx_1\\
&=\int_{-\ell/2}^{\ell/2} \int_{-\ell/2}^{\ell/2} B e^{-|x_2-x_1|/{\ell_0}}dx_2 dx_1\\
&=2B \int_{-\ell/2}^{\ell/2} \int_{x_1}^{\ell/2}  e^{-(x_2-x_1)/{\ell_0}} dx_2 dx_1\\
&=2B \ell_0^2 \left( \frac{\ell}{\ell_0} - 1 + e^{-\ell/{\ell_0}} \right).\\
\end{split}
\end{equation}
Also, 
\begin{equation} 
\begin{split}
&\int_{\ell/2}^{\mathcal{L}/2} \int_{-\ell/2}^{\ell/2} \left( \langle j(x_1)j(x_2) \rangle - \langle j(x_1) \rangle\langle j(x_2) \rangle \right) dx_2 dx_1\\
&=\int_{\ell/2}^{\mathcal{L}/2} \int_{-\ell/2}^{\ell/2}  B e^{-|x_2-x_1|/{\ell_0}}dx_2 dx_1\\
&=B\int_{\ell/2}^{\mathcal{L}/2}  e^{-x_1/\ell_0}dx_1   \int_{-\ell/2}^{\ell/2} e^{x_2/\ell_0}dx_2 \\
&=B \ell_0^2 \left( -e^{(\ell - \mathcal{L})/(2\ell_0)}+1+e^{-(\mathcal{L}+\ell)/(2\ell_0)}-e^{-\ell/\ell_0}\right)\\
\end{split}
\end{equation}
and taking $\lim{\mathcal{L} \to \infty}$  gives,
\begin{equation} 
\begin{split}
&\int_{\ell/2}^{\mathcal{L}/2} \int_{-\ell/2}^{\ell/2} \left( \langle j(x_1)j(x_2) \rangle - \langle j(x_1) \rangle\langle j(x_2) \rangle \right) dx_2 dx_1\\
&=B \ell_0^2 \left( 1-e^{-\ell/\ell_0}\right).
\end{split}
\end{equation}
Substituting into \ref{alphaflux_sup1} gives,
\begin{equation}
\begin{split}
&\alpha=\frac {2B \ell_0^2 \left( 1-e^{-\ell/\ell_0}\right)}{2B \ell_0^2 \left( \frac{\ell}{\ell_0} - 1 + e^{-\ell/{\ell_0}} \right)}\\
&=\dfrac{1-e^{-\ell/\ell_0}} { \frac{\ell}{\ell_0} - 1 + e^{-\ell/{\ell_0}}}.
\end{split}
\end{equation}
So, we obtain
\begin{equation} \label{LFT_final}
\ln \left( \frac{\mathrm{p}(\Omega_{\ell,t} =A)}{\mathrm{p}(\Omega_{\ell,t} =-A)} \right)= (1+\dfrac{1}{\frac{\ell}{\ell_0 (1-e^{- \ell/ \ell_0})}.                   -1})A
\end{equation}

Now, we consider equation (12) of the letter, which is valid for equilibrium states: 
\begin{equation}
\int_0^\mathcal{L}  \langle j_{t}(0) j_{t}(x) \rangle \mathrm{d}x = \dfrac{\langle (\Delta x)^2 \rangle \sum_{i=1}^N c_i^2}{2\mathcal{L}}
\end{equation}
To derive this equation, the quantity $\langle (J_t)^2 \rangle$ is computed by two methods.  First we note that 

\begin{equation}
\begin{split}
\langle (J_t)^2 \rangle &= \langle (\int_0^t \sum_{i=1}^N c_i v_{x,i}(s) \mathrm{d}s)^2 \rangle        \\ 
                        &= \langle (\sum_{i=1}^N c_i \Delta x_{i})^2 \rangle       \\ 
                        &= \sum_{i=1}^N \sum_{j=1}^N c_i c_j \langle \Delta x_{i}\Delta x_{j} \rangle       \\ 
						&= \sum_{i=1}^N c_i^2 \langle (\Delta x_{i})^2 \rangle + \sum_{i=1}^N \sum_{j=1, j\neq i}^N c_i c_j \langle \Delta x_{i}\Delta x_{j} \rangle       \\ 
						\label{sums}
                        &= \langle (\Delta x)^2 \rangle  \sum_{i=1}^N c_i^2 + \langle \Delta x_{1}\Delta x_{2} \rangle \sum_{i=1}^N  \sum_{j=1, j\neq i}^N c_i c_j    \\
\end{split}
\end{equation}
where $\Delta x_i = \int_0^t v_{x,i}(s) \mathrm{d}s$ is the $x$-displacement during a time $t$ of particle $i$ and we have assumed that particles are indistinguishable so $\langle \Delta x_i \rangle=\langle \Delta x \rangle$ for all $i$ and $\langle \Delta x_i \Delta x_j\rangle=\langle \Delta x_1 \Delta x_2\rangle$ for all $i$ and $j$. Note that these averages refer to an equilibrium state, {\it i.e.} when there is no external field. In the large system limit, the second term can be ignored because the region in which there is correlation between the motion of particles is finite.  Therefore the first equation becomes:
\begin{equation}
\langle (J_t)^2 \rangle = \langle (\Delta x)^2 \rangle  \sum_{i=1}^N c_i^2 
\end{equation}

On the other hand, the quantity $\langle (J_t)^2 \rangle$ can also be computed via the flux density:
\begin{equation}
\begin{split}
&\langle (J_t)^2 \rangle = \langle{ ( \int_{-\mathcal{L}/2}^{\mathcal{L}/2} j_{t}(x) dx ) ^2}  \rangle\\
&= \int_{-\mathcal{L}/2}^{\mathcal{L}/2} \int_{-\mathcal{L}/2}^{\mathcal{L}/2} \langle j_t(x_1)j_t(x_2) \rangle dx_2dx_1  \\
&= 2\int_{-\mathcal{L}/2}^{\mathcal{L}/2} \int_{x_1}^{\mathcal{L}/2} \langle j_t(x_1)j_t(x_2) \rangle dx_2dx_1.  \\
\end{split}
\end{equation}

If we assume that the system is homogeneous on average, so $\langle j_t(x_1)j_t(x_2) \rangle= \langle j_t(0)j_t(x_2-x_1) \rangle$, then:

\begin{equation}
\begin{split}
\langle (J_t)^2 \rangle = 2\int_{-\mathcal{L}/2}^{\mathcal{L}/2} \int_{x_1}^{\mathcal{L}/2}  \langle j_t(0)j_t(x_2-x_1) \rangle dx_2dx_1.  \\
\end{split}
\end{equation}

Now making a change of variables $a_1=(x_2-x_1)$, $a_2=(x_1+x_2+\mathcal{L})$ where $dx_1dx_2=da_1da_2/2$ we obtain,
\begin{equation}
\begin{split}
&\langle (J_t)^2 \rangle =\int_{0}^\mathcal{L}  \int_{a_1}^{2\mathcal{L}-a_1} \langle j_t(0)j_t(a_1) \rangle da_2da_1  \\
&= 2 \int_{0}^\mathcal{L} (\mathcal{L}-a_1)\langle j_t(0)j_t(a_1) \rangle da_1  \\
&= 2 \int_{0}^\mathcal{L} (\mathcal{L}-x)\langle j_t(0)j_t(x) \rangle dx  \\
\end{split}
\end{equation}
For a periodic system, $\langle j_t(0)j_t(x) \rangle=\langle j_t(0)j_t(\mathcal{L}-x) \rangle$, so we can write
\begin{equation}
\begin{split}
\langle (J_t)^2 \rangle &= 2( \int_{0}^{\mathcal{L}/2} (\mathcal{L}-x)\langle j_t(0)j_t(x) \rangle dx \\
&+\int_{\mathcal{L}/2}^\mathcal{L} (\mathcal{L}-x)\langle j_t(0)j_t(\mathcal{L}-x) \rangle dx) \\
&= 2( \int_{0}^{\mathcal{L}/2} (\mathcal{L}-x)\langle j_t(0)j_t(x) \rangle dx \\
&+\int_{0}^{\mathcal{L}/2} x \langle j_t(0)j_t(x) \rangle dx)   \\
&= 2\mathcal{L} \int_{0}^{\mathcal{L}/2} \langle j_t(0)j_t(x) \rangle dx \\
\end{split}
\end{equation}

Thetwo relationships for $\langle (J_t)^2 \rangle$ can be combined to give,
\begin{equation}
\int_0^{\mathcal{L}/2}  \langle j_{t}(0) j_{t}(x) \rangle \mathrm{d}x = \dfrac{\langle (\Delta x)^2 \rangle \sum_{i=1}^N c_i^2}{2\mathcal{L}}.
\label{numC}
\end{equation}

For Brownian particles it is expected that,
\begin{equation}
\langle j_{t}(0) j_{t}(0) \rangle \sim \dfrac{ \langle \mid \Delta x \mid \rangle \sum_{i=1}^N c_i^2}{\mathcal{L}}.
\label{denC}
\end{equation}

Substituting eqns \ref{numC} and \ref{denC} into eqns \ref{defL} and \ref{l0}, and taking the low field limit so $\langle j_t(0) \rangle \to 0$ gives,

\begin{equation} \label{ell0final_sup}
\ell_0 = \dfrac{\langle (\Delta x)^2 \rangle }{2 \langle \mid \Delta x \mid \rangle} 
\end{equation}

Assuming that the distribution of $\Delta x$ is Gaussian  ($\mathrm{p}(\Delta x=A)=\frac{1}{\sqrt(2 \pi \langle (\Delta x)^2 \rangle} e^{ - A^2/(2\langle (\Delta x)^2\rangle)}$) so $\langle |\Delta x| \rangle = 2\int_0^\infty  A \mathrm{p}(\Delta x=A) dA=\sqrt{2\langle (\Delta x)^2\rangle)/\pi}$, gives:

\begin{equation} \label{ell0final_sup}
\ell_0 \sim \sqrt{ \dfrac{\pi \langle (\Delta x)^2 \rangle}{8} }.
\end{equation}

This is the result given as equation (14) in the letter.